\newcommand{\CLASSINPUTtoptextmargin}{0.87in}
\newcommand{\CLASSINPUToutersidemargin}{0.71in}

\documentclass[conference,letterpaper]{IEEEtran}
\IEEEoverridecommandlockouts
\usepackage{cite}
\usepackage{graphicx}
\usepackage{array}
\usepackage{caption}
\usepackage{subcaption}
\usepackage{makecell}
\usepackage{gensymb}
\usepackage{amsmath,amssymb,amsfonts}

\input{accs}

\usepackage{etoolbox}
\usepackage{nomencl}
\makenomenclature

\providetoggle{nomsort}
\settoggle{nomsort}{true} 

\makeatletter
\iftoggle{nomsort}{%
    \let\old@@@nomenclature=\@@@nomenclature        
        \newcounter{@nomcount} \setcounter{@nomcount}{0}%
        \renewcommand\the@nomcount{\two@digits{\value{@nomcount}}}
        \def\@@@nomenclature[#1]#2#3{
          \addtocounter{@nomcount}{1}%
        \def\@tempa{#2}\def\@tempb{#3}%
          \protected@write\@nomenclaturefile{}%
          {\string\nomenclatureentry{\the@nomcount\nom@verb\@tempa @[{\nom@verb\@tempa}]%
          \begingroup\nom@verb\@tempb\protect\nomeqref{\theequation}%
          |nompageref}{\thepage}}%
          \endgroup
          \@esphack}%
      }{}
\makeatother

\usepackage{graphicx}
\usepackage{textcomp}
\usepackage{xcolor}
\def\BibTeX{{\rm B\kern-.05em{\sc i\kern-.025em b}\kern-.08em
    T\kern-.1667em\lower.7ex\hbox{E}\kern-.125emX}}

\PassOptionsToPackage{hyphens}{url}
\usepackage{url}
\begin{document}

\title{Smart Data Mapping for Connecting Power System Model and Geospatial Data\\
}

\author{\IEEEauthorblockN{Xue Li\IEEEauthorrefmark{1}, Kishan Prudhvi Guddanti\IEEEauthorrefmark{2}, Samrat Acharya\IEEEauthorrefmark{2}, Patrick Royer\IEEEauthorrefmark{1}, Xiaoyuan Fan\IEEEauthorrefmark{2}, Marcelo Elizondo\IEEEauthorrefmark{2} 
\IEEEauthorblockA{\IEEEauthorrefmark{1}Earth Systems Science Division, Pacific Northwest National Laboratory, Richland, WA, USA\\
\IEEEauthorrefmark{2} Electricity Infrastructure \& Building Division, Pacific Northwest National Laboratory, Richland, WA, USA\\
Emails: \{xue.li$|$kishan.g$|$samrat.acharya$|$patrick.royer$|$xiaoyuan.fan$|$marcelo.elizondo\}@pnnl.gov}
}}

\maketitle

\begin{abstract}
Knowing the geospatial locations of power system model elements and linking load models with end users and their communities are the foundation for analyzing system resilience and vulnerability to natural hazards. However, power system models and geospatial data for power grid assets are often developed asynchronously without close coordination. Creating a direct mapping between the two is a challenging task, mainly due to heterogeneous data structures, target uses, historical legacies, and human errors. This work aims to build an automatic data mapping workflow to connect the two, and to support energy grid resilience studies for Puerto Rico. The primary steps in this workflow include constructing graphs using geospatial data, and aligning them to the transmission networks defined in the power system data. The results have been evaluated against existing manual mapping practices for part of the Puerto Rico Power Grid model to illustrate the performance of such auto-mapping solutions.
\end{abstract}


\nomenclature[A]{\(PR\)}{\textcolor{teal}{Puerto Rico.}}
\nomenclature[A]{\(SITE\)}{\textcolor{teal}{Location/node from the geospatial data.}}
\nomenclature[A]{\(BUS\)}{\textcolor{teal}{Node/bus from the power system model data.}}
\nomenclature[A]{\(SITE-BUS\ mapping\)}{\textcolor{teal}{Indicates the mapping to correlate the SITE and BUS information.}}


\section{Introduction}

It is increasingly important to plan for power systems that are resilient to natural disasters and extreme weather conditions like hurricanes and flooding. A key aspect of resilience planning is to link datasets and models that have been traditionally not synchronized and developed separately \cite{b3}. Datasets are disconnected within power grid domain, e.g., planning and protection datasets; and the disconnection is larger between different domains, e.g., transmission power flow models are usually not linked with infrastructure assets information. Linking these datastes can help plan, design, and manage resilience activities supporting emergency response and preparedness. Geospatial mapping of power system data is critical as it can provide power system planners, investors, operators, and researchers with timely information and cross-domain associations.

The significance of accurate geospatial mapping is further underscored by the wide-spread yet devastating impact of natural hazards on Puerto Rico power grid. For instance, Hurricane Irma and Hurricane Maria were destructive storms that struck Puerto Rico two weeks apart in September 2017, causing extensive damage to approximately 80\% of Puerto Rico's power grid, primarily impacting transmission and distribution lines \cite{nypa}. This resulted in the most significant power outage in the history of the United States, with a lengthy recovery period of 328 days required for full restoration \cite{kwasinski2019hurricane}. Another hurricane, Fiona, made landfall in Puerto Rico on September 14, 2022, causing a blackout across the entire island \cite{luma}. The restoration process lasted nearly a month, and approximately 20\% of customers experienced power loss for over 10 days. All highlight the urgent needs of all available tools to support grid operators in power grids planning, operation, and restorations.

Recognizing the need for mitigating and overcoming such devastating impacts, a risk-based resilience planning and contingency analysis was proposed for Puerto Rico's power grid \cite{b2}, with a specific emphasis on the effects of hurricanes. The framework leveraged advanced tools such as the Dynamic Contingency Analysis Tool (DCAT) that simulate grid cascading failures from extreme events and the Electrical Grid Resilience and Assessment System (EGRASS) Tool that estimates the sequence of infrastructure failures to be simulated with DCAT. This framework highlighted the critical role of geospatial mapping of power system data especially for resilience studies \cite{b3}.

Several recent studies have utilized geospatial data to map power system topology \cite{us_eia, openstreet,arderne2020predictive, ml_worldbank}. OpenStreetMap and the U.S. Energy Information Administration provide publicly accessible data for global power networks and U.S. transmission-level power network mapped with geospatial information \cite{openstreet, us_eia}. Arderne \textit{et al.} \cite{arderne2020predictive} have mapped global transmission and distribution lines using open access data on electricity and road networks, satellite imagery, and socio-economic data such as population, electrification rates, night-time lights, and land cover. The authors in \cite{ml_worldbank} have used a convolutional neural network to estimate high-voltage transmission infrastructure in Nigeria, Pakistan, and Zambia. However, the accuracy of the global and regional mapping of power networks with geospatial data is still limited, requiring a more focused study for a particular region. 

To better assist grid transformation and renewable integration in Puerto Rico and enhance grid resilience, this study focuses on the geospatial mapping of high- and medium-voltage transmission-level power infrastructure in Puerto Rico.



\section{Power System and Geospatial Data}
\subsection{Power System Model Data}
\label{subsec:psse_data}

{In this work, the smart data mapping was applied to the transmission system network of Puerto Rico power grid. This sytem has $1367$ buses, out of which $17, 115,$ and $ 1245$ buses are at $230$kV, $115$kV, and $\leq 38$kV respectively. The grid has $1516$ transmission lines spanning over $2500$ miles; about $1200$ miles of the transmission lines are at transmission level ($\geq 115$kV), and about $1300$ miles of transmission line network are at subtransmission level ($\leq 38$kV), which typically tends to result in a more mesh-like structure at some locations. The grid also has more than $150$ two winding transformers, which are distributed among $115$kV, $40$kV, and $\leq 38$kV ($24$, $37$, and $94$ respectively). There are also $26$ three winding transformers.}

\subsection{Geospatial Data}
\label{subsec:gis_data}

This study was built upon geospatial datasets with a great level of details provided by utility partners, though such information has been anonymized in this paper to ensure proper data protection. In general, datasets most relevant to transmission asset mapping include site and transmission line; {the site information from geospatial dataset includes building footprint polygons of power plants, transmission centers, switch yards, substations, and others}, the embedded attributes include name, type, and voltage range. Transmission lines are line segments with attributes including voltage level, circuit ID, and line topology. There are no joinable fields between transmission lines and sites.



Despite great detail of asset locations and attributes, building a graph (nodes and edges) with the geospatial datasets remains challenging given the lack of built-in data connections and various data quality issues. For example, the topology-related fields in the transmission line segments contain errors and missing data that prevent successful graph building with those fields. {Therefore, the connections between line segments or buildings and lines are more attainable through geospatial reasoning based on the closeness of data objects.} However, this solution is still prone to locational and thematical errors which will be elaborated in the data preprocessing section (Section~\ref{subsec:geopreprocess}).

\subsection{Data Connections and Discrepancies}
\label{subsec:diffs}

In general, we expect some mapping relationship between \textbf{buses (from power system model data)} and \textbf{sites (from geospatial data)}, and between power system lines and geospatial transmission lines. In many cases, we found 1:1 mapping between sites and buses usually through their name fields. {Additionally, the site data can be geospatially joined with a planning zone polygon data to add a "zone" field that is relevant to the "area" field found in the bus data. Although there is no common field exists in the two line datasets for direct mapping, it is 
attainable through matching two graphs using bus-site mapping information.} However, there are many data discrepancies that complicate the mapping process.

Between buses and sites, the name fields are seldom matching exactly for a straightforward mapping process between buses and sites. Besides such inconsistent naming, bus-site mapping is usually not one on one but many-to-many. For example, a bus may be represented by several sites including substations and a switch yard. And a transmission center site is always related to multiple buses at different voltage levels, usually connected through transformers. Moreover, not all buses have a site representation, and vice versa. One example is line tap which usually does not have a site; another example is legacy bus that no longer has physical sites.

Different topology is expected between the power system and geospatial graphs, even by assuming perfect graph building from the geospatial dataset. This is due to the fact that power system data is updated more frequently than geospatial data, so there is usually a gap in the system versions. Therefore, power system lines may not find a counterpart in geospatial transmission lines, or vice versa.

Therefore, a smart mapping approach is needed to navigate through all kinds of data discrepancies and make connections between power system elements to geospatial assets.


\section{The Methodology}
\label{sec:methodology}
\subsection{Geospatial Data Preprocessing}
\label{subsec:geopreprocess}

The purpose of geospatial data preprocessing is to build graphs from the site and transmission line datasets. The first step is to merge some sites into groups, if they are likely represented by single buses in the power system model. Next, the transmission line dataset needs to go through some initial cleanings to fix data issues before being used to build edges connecting groups of sites. The outputs of preprocessing are geo-graphs at different voltage levels, with site groups as nodes and transmission lines as edges.

\begin{figure}[htp]
    \centering    
    \begin{subfigure}[b]{0.34\textwidth}
        \centering
        \includegraphics[width=\textwidth]{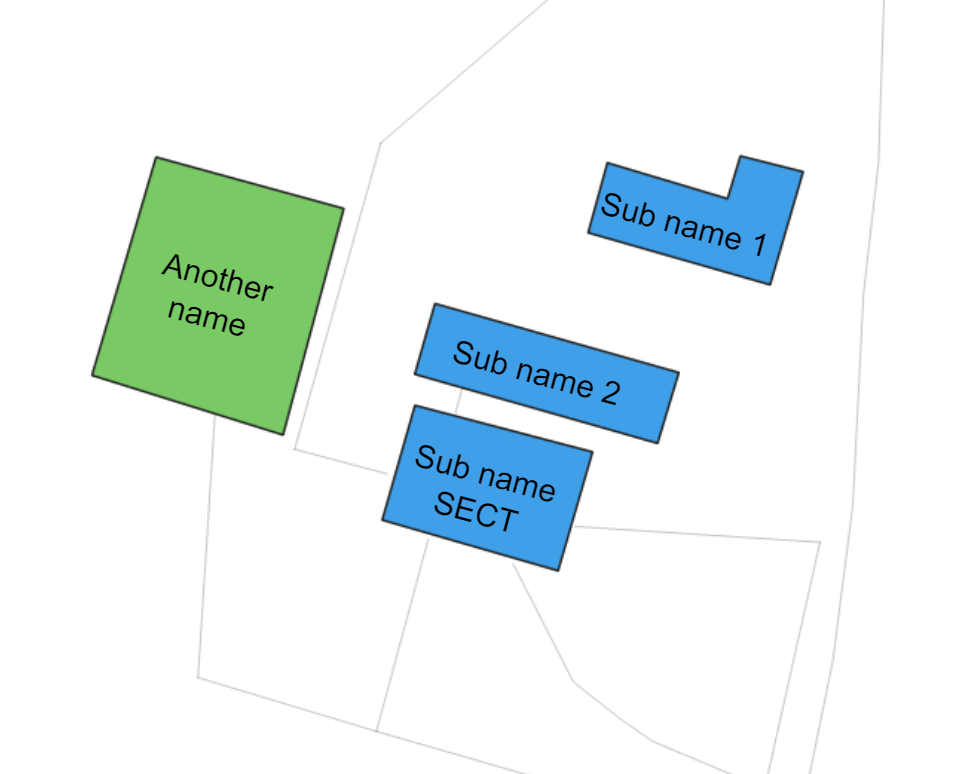}
        \caption{Site grouping}
        \label{fig:site_grouping}
     \end{subfigure}
     \begin{subfigure}[b]{0.14\textwidth}
        \centering
        \includegraphics[width=.7\textwidth]{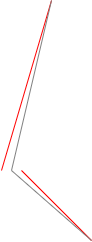}
        \caption{Line geometry repairing}
        \label{fig:line_snapping}
     \end{subfigure}
    \caption{Illustration of geospatial data preprocessing}
    \label{fig:line_snapping}
\end{figure}

A common scenario of site grouping involves a transmission center or switch yard and its adjacent substations (Figure \ref{fig:site_grouping}). Two major clues for potential grouping are closeness in names and locations. A multi-criteria clustering was applied to sites in each planning zones. The closeness of names was calculated based on the Longest Common Substring (LCS) Distance implemented in the stringdist library in R \cite{RJ-2014-011}, and the geospatial distances between sites were calculated using the sf library in R \cite{sf}. Hierarchical clustering of sites was conducted with name and locational distances separately, and the grouping labels were combined to form new groups if sites were considered together with both criteria. The name strings were concatenated and the voltage ranges were expanded according to the sites included in each group.

Transmission line data cleaning included two steps. First is fixing missing circuit IDs which may be in the form of "NONE", "0", or simply no data. The logic is to find the circuit ID of its immediate neighbor for every segment with missing IDs and do so repeatedly until no more segments can be fixed with this method. The second step of fixing involves snapping line segment vertices to their neighboring vertices if they are not exactly collocated (Figure \ref{fig:line_snapping}). This was realized through point clustering based on the Density-based Spatial Clustering of Applications with Noise (DBSCAN) \cite{dbscan}. This algorithm was applied to line segments by each circuit ID to avoid having overlapping lines which may introduce undesired clustering and thus wrong connection of lines on different circuits.


Finally, the sfnetwork package in R \cite{sfnetwork} was used to build geospatial graphs that connect grouped sites through cleaned transmission lines. An initial graph can be built with all the line segments from each particular circuit at a certain voltage level. For each circuit, a subset of site groups was considered connected if they fell within a distance threshold of the circuit lines and had the line voltage within their ranges. When there are two site groups on a circuit, a simple geospatial graph is built with two nodes and an edge following the geospatial line. In the case of multiple nodes, a complete geospatial graph is built connecting all pairs of nodes through geospatial transmission lines and using line length as weight. Then prune the complete graph to its minimum spanning tree and add back pruned edges, ranked from shortest to longest, if they represent a shorter path between the two nodes than that through the current graph.

\subsection{Graph Mapping}
\label{subsec:smartmapping}


Mapping between these two related but not isomorphic graphs involves a combination of different strategies (Algorithm \ref{alg:algo1}). 
Fuzzy name matching {(Algorithm \ref{alg:algo1}, line $7$)} was first conducted at each voltage level and results with high confidence were used as seeds ({initialization solution}) to search additional mappings through combined name- and topology-based mapping {(Algorithm \ref{alg:algo1}, line $15$)}. {To solve complicated cases that are typically situated at lower kV level mapping, the proposed mapping was approached from higher to lower voltage levels while passing down the set of seeds inherited from higher voltage levels through transformers and transmission centers {(Algorithm \ref{alg:algo1}, line $12$)}.}
In the end, power system branches were joined with geospatial lines using the outputs of the proposed bus-site mapping algorithm. More details of the major steps are explained below.

The first clue for connecting buses and site groups is through names. However, the name fields are seldom matching exactly for direct table joining. Some of the interfering factors include voltage in one but not the other (e.g., CANA 115 vs. CANA), inconsistent uses of abbreviations (e.g., SJSP vs. San Juan SP), non-unique names (e.g., COSTCO), not to mention cases when two names appear completely irrelevant. Therefore, fuzzy name matching was first conducted to identify potential links with string similarities defined by the LCS method. Similarity of the area field was used to prevent wrong assignments of mapping across different areas, especially in the case of non-unique names (Algorithm \ref{alg:algo2}).

Using high-confidence name-matching results and inherited seeds (if not at 230kV), a topology-based mapping strategy was implemented to grow mapping from seeds through their neighborhoods (Algorithm \ref{alg:algo3}). Name-based matching was applied again at neighborhood level aiming at a unique assignment of site for each bus. The seed itself was considered in the neighborhood mapping to provide potential solutions for non-site buses and non-bus sites.


\begin{algorithm}[ht]
\caption{{Smart Graph Mapping Approach.}}\label{alg:algo1}
\begin{algorithmic}[1]
 \Statex \textbf{Input:} 
 \LState GIS graph data: $\mathcal{G}^{gis}$ where $\mathcal{G}^{gis}$ = $\left(\mathcal{\textbf{V}}^{gis}, \mathcal{E}^{gis}\right)$
 \LState Power system model (PSM) graph data: $\mathcal{G}^{ps}$ where $\mathcal{G}^{ps}$ = $\left(\mathcal{\textbf{V}}^{ps}, \mathcal{E}^{ps}\right)$
 \LState GIS node attribute information: $\mathcal{\textbf{V}}^{gis} \in \mathbb{R}^{(n,4)}$ where $n$ = number of sites; $\mathcal{\textbf{V}}^{gis}$ =  $\left[\text{id},\text{kV},\text{name}, \text{zone}\right]_{i}$ $\forall i \in \left[1,n\right]$ 
 \LState PSM node attribute info.: $\mathcal{\textbf{V}}^{ps} \in \mathbb{R}^{(m,4)}$ where $m$ = number of buses; $\mathcal{\textbf{V}}^{ps}$ =  $\left[\text{id},\text{kV},\text{name}, \text{area}\right]_{i}$ $\forall i \in \left[1, m\right]$ 

 \Statex \textbf{Output:} PSM graph data with \underline{GIS site mapping}: $\mathcal{\textbf{V}}^{ps}_{final}\in \mathbb{R}^{(m,5)}$ where $m$ = number of buses and $\mathcal{\textbf{V}}^{ps}_{final}$ = $\left[\text{id},\text{kV},\text{name}, \text{area}, \text{\textbf{mapped site id}}\right]_{i}$ $\forall i \in \left[1, m\right]$.
 
 \LState $\mathcal{\textbf{V}}^{ps}_{final}$ = $\mathcal{\textbf{V}}^{ps}$ $+\ zeros\left(m,1\right)$;
 \For{$i =\{230, 115, 38\}$} \Comment{}{For different kV levels.}
 \LState $\mathcal{\textbf{M}}$ = \textit{\textbf{nameMatching}}$\left(\mathcal{G}^{gis}, \mathcal{G}^{ps}\right)$; \Comment{}{Perform initial bus-site mapping through name matching.}
  \LState $\mathcal{\textbf{M}}$ = $\mathcal{\textbf{M}}$(similarity score$\geq 0.6$); 
 \Comment{}{Retain bus-site mappings that have high similarity scores as seeds.} 
 \If{i == 230}
 \LState $\mathcal{\textbf{V}}^{ps}_{final}(:,5)$ = $\mathcal{\textbf{M}}$; \Comment{}{Initialize bus-site mappings with name-based seeds.}
 \Else{}
 \LState $\mathcal{\textbf{V}}^{ps}_{final}$ = $inheriteSeeds\left(\mathcal{\textbf{V}}^{ps}_{final}, \mathcal{G}^{ps}\right)$
 \Comment{}{Pass down previous bus-site mapping through transformers.}
 \LState $\mathcal{\textbf{V}}^{ps}_{final}(:, 5)$ = $merge\left(\mathcal{\textbf{V}}^{ps}_{final}(:, 5), \mathcal{\textbf{M}}\right)$
 \Comment{}{Merge previous iteration's saved mappings and current seeds.}
 \EndIf
 \LState $\mathcal{\textbf{V}}^{ps}_{final}$ = \textit{\textbf{topoMatching}}$\left(\mathcal{G}^{gis}, \mathcal{G}^{ps}, \mathcal{\textbf{V}}^{ps}_{final}\right)$ \Comment{}{Grow site-bus mapping through topology neighborhood matching to fill remaining gaps.}

\LState Branches are mapped using the node mappings from previous steps.

    \EndFor
  
  \State \Return $\mathcal{\textbf{V}}^{ps}_{final}$;
\end{algorithmic}
\end{algorithm}

\subsection{Performance Evaluation}

Performance of asset mapping depends on both the quality of geospatial graph and the graph mapping algorithms and strategies. Graph built from imperfect geospatial datasets lays the basis for graph mapping and also defines the ceiling of mapping performance. The quality of geospatial graph can be measured by the percentage of sites and transmission lines being built into the final graph. Graph mapping performance was evaluated through both mapping completeness and validation against manual mapping results. All high-voltage (230/115kV) assets and a small number of low-voltage assets (38kV) were previously examined with manual efforts, which were used to validate the smart mapping results.

\begin{algorithm}[ht]
\caption{{BUS-SITE mapping through name matching.}}\label{alg:algo2}
\begin{algorithmic}[1]
  \Statex \textbf{Function:} $\mathcal{\textbf{M}}$ = \textit{\textbf{nameMatching}}$\left(\mathcal{G}^{gis}, \mathcal{G}^{ps}\right)$
 \Statex \textbf{Input:} 
 \LState GIS graph data: $\mathcal{G}^{gis}$ and PSM graph data: $\mathcal{G}^{ps}$

 \Statex \textbf{Output:} Vector that contains a mapped site location (from $\mathcal{G}^{gis}$) for each bus (in $\mathcal{G}^{ps}$) = $\mathcal{\textbf{M}} \in \mathbb{R}^{(m,1)}$. $m$ is the number of buses in power system model data.
 
 \Statex \underline{\textbf{Approach:}}
 \LState Compute $\mathcal{\textbf{S}}^{(b.n,s.n)} \in \mathbb{R}^{(m,n)}$ \Comment{}{Similarity score (LCS) matrix for \textit{bus names} (b.n) and \textit{site names} (s.n)}
 \LState Compute $\mathcal{\textbf{S}}^{(a.n,z.n)} \in \mathbb{R}^{(m,n)}$ \Comment{}{Similarity score (LCS) matrix for \textit{area names} (a.n) and \textit{zone names} (z.n)}
 \LState $\mathcal{\textbf{S}}^{merge}$=$\left\{x| x = min(e_1,e_2)\right\}$ where $e_1 \in \mathcal{\textbf{S}}^{(b.n,s.n)}$ and $e_2 \in \mathcal{\textbf{S}}^{(a.n,z.n)}$. \Comment{}{Merge $\mathcal{\textbf{S}}^{(b.n,s.n)}$ and $\mathcal{\textbf{S}}^{(a.n,z.n)}$ with worse similarity scores.}
 \LState $\mathcal{\textbf{M}}$ = $argmax\left(\mathcal{\textbf{S}}^{merge}\right) \in \mathbb{R}^{(m,1)}$;
 \Comment{}{Select the site from $\mathcal{\textbf{V}}^{gis}$ that has highest similarity score (among other sites) for each node in $\mathcal{\textbf{V}}^{ps}$ i.e., \underline{bus-site mapping}.}
  
  \State \Return $\mathcal{\textbf{M}}$;
\end{algorithmic}
\end{algorithm}
\begin{algorithm}
\caption{{BUS-SITE mapping through topology matching.}}\label{alg:algo3}
\begin{algorithmic}[1]
  \Statex \textbf{Function:} $\mathcal{\textbf{M}}$ = \textit{\textbf{topoMatching}}$\left(\mathcal{G}^{gis}, \mathcal{G}^{ps}, \mathcal{\textbf{V}}^{ps}\right)$

 
 \Statex \underline{\textbf{Approach:}}
 \LState $\mathcal{\textbf{V}}^{ps}_{done}$ = $\mathcal{\textbf{V}}^{ps}$
 \LState $\mathcal{\textbf{L}}$ = empty table \Comment{}{Table stores candidates for mapping}
 \LState count = 0;  $\mathcal{\textbf{L}}_{prev}$ = random valued table;
 \While{count$<2000$ \text{\textbf{AND}} $\mathcal{\textbf{L}}_{prev}$ $\neq$ $\mathcal{\textbf{L}}$}  
 \LState $\mathcal{\textbf{L}}_{prev}$ = $\mathcal{\textbf{L}}$;
 \LState count = count + 1;
  \For{$i =\{1, 2, \cdots, \textit{rows}\left(\mathcal{\textbf{V}}^{ps}(:,5)\right)\}$} \Comment{}{For every available mapped bus via initial input or new solutions from {\textit{\textbf{candidateConfirm}}} (line 29), expand its neighbors and map any neighbors that are not mapped yet.}
 \LState Selected bus ($b$) = $\mathcal{\textbf{V}}^{ps}(i,5)$;
 \LState $b_{\mathcal{N}(gis)}$ = $\mathcal{N}(b)\left(\mathcal{G}^{gis}\right)$; \Comment{}{Neighbors of $b$ in $\mathcal{G}^{gis}$.}
 \LState $b_{\mathcal{N}(ps)}$ = $\mathcal{N}(b)\left(\mathcal{G}^{ps}\right)$; \Comment{}{Neighbors $b$ in $\mathcal{G}^{ps}$.}
 \If{(neighbors exist for $b$ in $\mathcal{G}^{gis}$ \textbf{AND} $\mathcal{G}^{ps}$)}
 \LState $\mathcal{\textbf{M}}$ = \textit{\textbf{nameMatching}}$\left(b_{\mathcal{N}(gis)}, b_{\mathcal{N}(ps)}\right)$; 
 \If{$m < n$} \Comment{}{Total buses $<$ total sites.}
 \LState flag = $True$ \Comment{}{If at least one site is mapped to multiple buses, otherwise its $False$} 
 \Else{}
 \LState flag = $False$ \Comment{}{If no sites are not mapped, otherwise its $True$}
 \EndIf
 \LState counter = 0
 \While{flag \textbf{AND} counter$<10$}
 \LState counter = counter + 1;
 \LState $\mathcal{\textbf{M}}$ = $updateDuplicates\left(b_{\mathcal{N}(gis)}, b_{\mathcal{N}(ps)},\right.$
 \Statex $\left. \mathcal{\textbf{M}}\right)$ \Comment{}{For buses mapped to same site, keep only the mapping with highest score and remap other buses to their next best site solutions.}
 \LState \textit{Update} flag variable using logic from steps $10$ through $16$.
 \EndWhile
 \Else{}
 \LState \textit{next}
 \EndIf
  \LState $\mathcal{\textbf{L}}$ = $merge(\mathcal{\textbf{L}},\mathcal{\textbf{M}})$ \Comment{}{Add potential mapping solutions from each seed to a candidate pool.}
 \EndFor
 \LState $\mathcal{\textbf{V}}^{ps}$ = $candidateConfirm(\mathcal{\textbf{L}})$ \Comment{}{Candidates can be confirmed by score ranking or multiple check-ins through different seeds regardless of scores.}
 \LState $\mathcal{\textbf{V}}^{ps}_{done}$ = $merge(\mathcal{\textbf{V}}^{ps}_{done},\mathcal{\textbf{V}}^{ps})$
 \EndWhile
  
  \State \Return $\mathcal{\textbf{V}}^{ps}_{done}$;
\end{algorithmic}
\end{algorithm}
\vspace{-1em}

\section{Results}
\subsection{Geospatial Graph Creation}
After site grouping and transmission line data cleaning, geospatial graphs were built for different voltage levels separately ({Section~\ref{subsec:geopreprocess}}). Table \ref{table:perf_graph} summarises the number of nodes and edges, the total length of transmission lines being built into the graphs, and percent coverage of the original dataset for each voltage level. At 230 kV, the geospatial graph building is largely complete. Lines not built in are short sections within transmission centers or long sections that do not appear to be connected through visual interpretation. At 115 kV, relatively more lines failed to be built/included into the graph due to remaining data issues not fully resolved in data preprocessing. Examples include large gaps between sites and lines and wrong labeling of line circuits that breaks connectivity. Graph building completeness further declined at 38kV level given the above-mentioned data inaccuracy. A unique network structure observed in the 38 kV geospatial data is parallel edges on single circuit, which cannot be resolved in the current algorithm. Despite all discrepancies, roughly 90\% of sites and transmission lines were successfully built into geospatial graphs, which laid the foundation for the subsequent graph mapping and ensures practical applications.

\subsection{Inter-Graph Mapping}
Smart mapping algorithms (Section~\ref{subsec:smartmapping}) were applied between power system and geospatial graphs from high to low voltages and achieved different levels of performances (Table \ref{table:vali}). At 230kV, all buses were successfully mapped to site groups including known legacy buses that were mapped arbitrarily to their neighbors. A manual check later confirmed the correctness of the mapping results. At 115kV level, majority of buses and sites were mapped with an overall accuracy of nearly 90\%. An initial mismatch in name-based seed generation (high name similarity but wrong mapping) misaligned two graphs locally and affected mapping correctness and completeness in its close neighborhoods. In addition, failed site grouping in a few places, due to dissimilar names, introduced more sites in the geospatial graph and complicated the graph mapping. Similar problems and less complete geospatial graphs further deteriorate the mapping performance at 38kV, but overall the smart mapping algorithms achieved decent performance with around 90\% buses and 80\% site groups mapped. Mapping accuracy was estimated at 72\% based on validation against previous manual mapping results over the San Juan area.
\vspace{-0.5em}
\begin{table}[!ht]
\begin{center}
\caption{Geospatial graph-building results}
\label{table:perf_graph}
\begin{tabular}{lllll} 
  \hline
   Voltage & \# site groups & \# lines & line length & \% coverage \\
  \hline
  230kV & 13 & 17 & 414 mi & 97.7\% \\
  115kV & 84 & 102 & 669 mi & 94.0\% \\
  38kV & 951 & 1126 & 1345 mi & 86.0\% \\
  \hline
\end{tabular}
\end{center}
\end{table}
\vspace{-2em}

\begin{table}[!ht]
\begin{center}
\caption{Graph mapping performance}
\label{table:vali}
\begin{tabular}{llll} 
  \hline
   Voltage & \% of buses mapped & \% of site groups mapped & accuracy \\
  \hline
  230kV & 100\% &100\%& 100\%\\
  115kV & 97.1\%& 93.0\% & 88.5\%\\
  38kV & 90.9\%& 78.9\%& 72.3\% \\
  \hline
\end{tabular}
\end{center}
\end{table}
\vspace{-1em}
\section{Discussions and Conclusions}
Both power system and geospatial datasets were available from utility company with a great level of detail for Puerto Rico. However, as also true in other systems, direct mapping between the two datasets was lacking and various data inconsistencies signifcantly burden such work. An automatic mapping workflow was described in this paper, which includes building graphs from imperfect geospatial datasets and matching them with power system graphs that are similar but not identical in both node name and network topology. Given the expected similarities and dissimilarities between the two sets of graphs, a combined mapping strategy was implemented including seed generation based on high scores in name similarity and topology-based mapping that grows seeds through their neighborhoods. Graph mapping was approached from the highest voltage level when the greatest similarity and simplicity were expected. Same mapping strategy was applied at lower voltage levels and with additional seeds inherited from transformers/transmission centers from higher levels. As a result, both geospatial graph-building and graph mapping achieved good performance, and certain degradation exists at lower voltage levels.

Having accurate mapping between geospatial information and power system models can enable power system planners and operators to make enriched and informed decisions for emergency preparedness and response. It can also help perform richer studies considering vulnerability under natural hazards for power system loads linked with end users and their communities. This type of analysis can be particularly important in areas impacted by natural hazards, where vulnerable communities may be disproportionately affected by power outages. With enhanced geospatial mapping of grid asset and community data, grid operator may implement and integrate such mapping in planning and operational practices, as a result, the improved disaster preparedness and situational awareness can help ensure the resilience and reliability of power systems, including weather-dependent renewable energy integration, natural disasters, extreme weather conditions, cyber-physical attacks, and more.

\section*{Acknowledgment}

The authors would like thank the Federal Emergency Management Agency (FEMA) and the U.S. Department of Energy (DOE) for funding this study, as well as our colleagues from Sandia National Laboratory and the engineering team from LUMA Energy. This study was conducted at Pacific Northwest National Laboratory, which is operated for the U.S. Department of Energy by Battelle Memorial Institute under Contract DE-AC05-76RL01830.

\bibliographystyle{IEEEtran}
\bibliography{ref}

\end{document}